# Phemenological Modelling of a Group of Eclipsing Binary Stars

Ivan L. Andronov[1, *], Mariia G. Tkachenko[1], Lidia L. Chinarova[2]

[1] Department "High and Applied Mathematics", Odessa National Maritime University, Odessa, Ukraine
[2] Astronomical Observatory, Odessa National University, Odessa, Ukraine

**Abstract**

Phenomenological modeling of variable stars allows determination of a set of the parameters, which are needed for classification in the "General Catalogue of Variable Stars" and similar catalogs. We apply a recent method NAV ("New Algol Variable") to eclipsing binary stars of different types. Although all periodic functions may be represented as Fourier series with an infinite number of coefficients, this is impossible for a finite number of the observations. Thus one may use a restricted Fourier series, i.e. a trigonometric polynomial (TP) of order $s$ either for fitting the light curve, or to make a periodogram analysis. However, the number of parameters needed drastically increases with decreasing width of minimum. In the NAV algorithm, the special shape of minimum is used, so the number of parameters is limited to 10 (if the period and initial epoch are fixed) or 12 (not fixed). We illustrate the NAV method by application to a recently discovered Algol-type eclipsing variable 2MASS J11080308-6145589 (in the field of previously known variable star RS Car) and compare results to that obtained using the TP fits. For this system, the statistically optimal number of parameters is 44, but the fit is still worse than that of the NAV fit. Application to the system GSC 3692-00624 argues that the NAV fit is better than the TP one even for the case of EW-type stars with much wider eclipses. Model parameters are listed.

**Keywords**

Astronomy, Stellar Astrophysics, Variable Stars, Eclipsing Binaries, Time Series Analysis, Data Reduction, Stars: Individual: 2MASS J11080308-6145589, GSC 3692-00624, RS Car

## 1. Introduction

Variability of stars indicates active processes in these objects and allows determination of physical parameters after complementary studies – photometrical, polarimetrical, spectral.

In the "General Catalogue of Variable Stars" (GCVS [33], available online at http://www.sai.msu.su/gcvs/gcvs/ ), there are listed 47968 stars (December, 2014), the variability of which was confirmed by independent studies. In the "Variable Stars Index" (VSX, http://aavso.org/vsx), there are listed more than 250 thousand stars, the majority of which were studied only once after discovery. Many stars are discovered as variables during huge photometric surveys (e.g. [24]).

There is a special set of the parameters needed for inclusion of the star to the GCVS or other catalogs of variable stars – co-ordinates; brightness range in stellar magnitudes; type of variability (currently officially declared in the GCVS [33] are ~70 types); for periodic stars – the period $P$ and the initial epoch $T_0$; asymmetry $M$-$m$ for the pulsating variables or the eclipse duration $D$ for eclipsing binary stars. Such parameters are called "phenomenological", contrary to the "physical" parameters like the masses, radiuses (or potentials), temperatures (or spectral classes) of the components of the binary system.

The physical parameters are typically determined by fitting observational light curve with theoretical ("physical") function and minimizing the test function (Eq. (1)). There are various programs for physical modeling (e.g. [11, 12, 13, 15, 29, 30, 41, 42) mostly based on the Wilson-Devinney algorithm [37,38,39,40]. The classical monographs on binary stars were presented by Kopal [16], Tsessevich [35] et al.

In the second part, we present expressions for the Least squares method, which is generalized for a case of generally correlated observational errors. These expressions are used for a study of a newly discovered Algol-type (EA) variable star 2MASS J11080308-6145589 [23].

The official classification of variable stars is presented in [33].

## 2. Mathematical Modeling

### 2.1. Generalized method of the least squares

#### 1. Linear model

The classical method of the least squares was described in many textbooks (e.g. [1,13,17,28]). It is based on minimization of the test function

$$\Phi_m = (\vec{x} - \vec{x_C}) \cdot (\vec{x} - \vec{x_C}), \quad (1)$$

where $\vec{x} = x_k$, $k = 1..n$ is the vector of observational values, $\vec{x_C} = x_{Ck}$ is the vector of computational values, which are dependent on some coefficients $C_\alpha$, $\alpha = 1..m$ and some argument $t$. Typically $t$ is scalar (e.g. time of observations), but may generally be a vector (e.g. co-ordinates).

The scalar product may be generally determined using the "metrical tensor":

$$\vec{a} \cdot \vec{b} = \sum_{i,k=1}^{n} h_{ik} a_i b_k, \quad (2)$$

but this expression is typically simplified to $h_{ik} = w_k \delta_{ik}$, where $\delta_{kk} = 1$, and $\delta_{ik} = 0$, if $i \neq k$, $w_k$ is "weight of the observation". The corresponding expression for the scalar product is

$$\vec{a} \cdot \vec{b} = \sum_{k=1}^{n} w_k a_k b_k, \quad (3)$$

Usually the weight is defined as $w_k = \frac{\sigma_0^2}{\sigma_k^2}$, where $\sigma_k$ is an accuracy of the $k^{th}$ observation, and a positive constant $\sigma_0$ is called "the unit weight error" (i.e. $w_k = 1$, if $\sigma_k = \sigma_0$). Most usual case is when the weights are set to $w_k = 1$ (i.e. assuming that the accuracy $\sigma_k$ of each value $x_k$ is constant, or neglecting possible small differences):

$$\vec{a} \cdot \vec{b} = \sum_{k=1}^{n} a_k b_k, \quad (4)$$

Next step for the method of the least squares is to approximate the signal with a linear combination of "basic" functions $f_\alpha(t)$ with corresponding coefficients $C_\alpha$:

$$x_C(t) = \sum_{\alpha=1}^{m} C_\alpha f_\alpha(t), \quad (5)$$

Obviously, one may write useful equations:

$$C_\alpha = \frac{\partial x_C}{\partial f_\alpha}, \quad (6)$$

$$f_\alpha = \frac{\partial x_C}{\partial C_\alpha}, \quad (7)$$

The conditions for the minimum of $\Phi_m$

$$\frac{\partial \Phi_m}{\partial C_\alpha} = -2(\vec{x} - \vec{x_C}) \vec{f_\alpha} = 0, \quad (8)$$

lead to the system of normal equations:

$$\sum_{\alpha=1}^{m} A_{\alpha\beta} C_\alpha = B_\beta, \quad (9)$$

where $A_{\alpha\beta} = \vec{f_\alpha} \vec{f_\beta}$, $B_\beta = \vec{x} \vec{f_\beta}$. The best fit coefficients are

$$C_\alpha = \sum_{\beta=1}^{m} A_{\alpha\beta}^{-1} B_\beta, \quad (10)$$

Here $A_{\alpha\beta}^{-1}$ is the matrix, which is inverse to the matrix of normal equations $A_{\alpha\beta}$.

#### 2. Statistical properties of coefficients and functions

The mathematical expectation of the deviations of the coefficients is $\langle \delta C_\alpha \rangle = 0$. The complete covariation matrix is

$$R_{\alpha\gamma} \stackrel{\text{def}}{=} \langle \delta C_\alpha \delta C_\gamma \rangle = \sum_{iqks=1}^{n} \mu_{iq} \sum_{\beta\varepsilon=1}^{m} A_{\alpha\beta}^{-1} A_{\gamma\varepsilon}^{-1} h_{ik} f_{\beta k} h_{qs} f_{\gamma s}, \quad (11)$$

and $\mu_{iq} \stackrel{\text{def}}{=} \langle \delta x_i \delta x_q \rangle$ is the covariation matrix of the deviations of observations. This is a statistically correct complete form for the matrix $R_{\alpha\gamma}$ for generally correlated deviations of observations, contrary to common simplification

$$R_{\alpha\gamma} = \sigma_{0m}^2 A_{\alpha\gamma}^{-1}, \qquad (12)$$

However, the complete form for the unbiased estimate of the squared unit weight error for the model with $m$ parameters is

$$\sigma_{0m}^2 = \frac{\Phi_m}{n-m} = \frac{1}{n-m} \sum_{i,k=1}^{n} h_{ik}(x_i - x_{Ci})(x_k - x_{Ck}) \qquad (13)$$

and may be simplified like in the Eq. (3) and (4). The simplification (Eq. 12) is valid also for a complete Eq. (2), if the matrixes $h_{ik}$ and $\mu_{ik}$ are related:

$$h_{ik} = \sigma_0^2 \mu_{ik}^{-1}, \qquad (14)$$

$$\mu_{ik} = \sigma_0^2 h_{ik}^{-1}, \qquad (15)$$

Otherwise one has to use a complete Eq. (11). Even more general case of the additional wavelet – like weights dependent on time and scale was described in [4].

The accuracy estimate of the coefficient $C_\alpha$ is $\sigma[C_\alpha] = \sqrt{R_{\alpha\alpha}}$.

The arbitrary function of coefficients $G(C_\alpha)$ may be expressed as the multivariate Taylor series for $C_\alpha = C_{0\alpha} + \delta C_\alpha$:

$$G(C_\alpha) = G(C_{0\alpha})$$
$$+ \sum_{\alpha=1}^{m} \frac{\partial G}{\partial C_\alpha} \delta C_\alpha + \frac{1}{2!} \sum_{\alpha,\beta=1}^{m} \frac{\partial^2 G}{\partial C_\alpha \partial C_\beta} \delta C_\alpha \delta C_\beta$$
$$+ \cdots + \frac{1}{N!} \sum_{\alpha_1,\ldots,\alpha_N=1}^{m} \frac{\partial^N G}{\partial C_{\alpha_1} \ldots \partial C_{\alpha_N}} \delta C_{\alpha_1} \ldots \delta C_{\alpha_N}$$
$$+ \cdots \qquad (16)$$

There are two important application of this expression. At first, one may compute the value of $G(C_\alpha)$ for "slightly other" values of the coefficients (e.g. after rounding their values). The mathematical expectation of $G(C_\alpha)$ assuming that the coefficients $C_\alpha = C_{0\alpha} + \delta C_\alpha$ are slightly different from the "true" values $C_{0\alpha}$ is

$$\langle G(C_\alpha) \rangle = G(C_{0\alpha}) + \frac{1}{2!} \sum_{\alpha,\beta=1}^{m} \frac{\partial^2 G}{\partial C_\alpha \partial C_\beta} R_{\alpha\beta} + \cdots \qquad (17)$$

Typically only the first term of this sum is taken into account, but the second one may be also important.

The error estimate $\sigma[G(C_\alpha)]$ may be determined from

$$\sigma^2[G(C_\alpha)] = \sum_{\alpha,\beta=1}^{m} R_{\alpha\beta} \frac{\partial G}{\partial C_\alpha} \frac{\partial G}{\partial C_\beta} + \cdots \qquad (18)$$

Particularly, Eq. (18) may be used for determination of the error estimate of the smoothing function (5)

$$\sigma^2[x_C(t)] = \sum_{\alpha,\beta=1}^{m} R_{\alpha\beta} f_\alpha(t) f_\beta(t) \qquad (19)$$

and its derivatives $x_C^{(q)}(t) \stackrel{\text{def}}{=} \partial^q x_C(t)/\partial t^q$ of degree $q$

$$\sigma^2[x_C^{(q)}(t)] = \sum_{\alpha,\beta=1}^{m} R_{\alpha\beta} f_\alpha^{(q)}(t) f_\beta^{(q)}(t) \qquad (20)$$

One may note that in a majority of publications the authors use abbreviated expression

$$\sigma^2[x_C(t)] = \sum_{\alpha=1}^{m} R_{\alpha\alpha} (f_\alpha(t))^2 = \sum_{\alpha=1}^{m} (\sigma[C_\alpha] f_\alpha(t))^2 \qquad (21)$$

The expressions (19) and (21) give same results *only* if the matrix $R_{\alpha\beta}$ is diagonal, i.e. $R_{\alpha\beta} = R_{\alpha\alpha} \delta_{\alpha\beta}$. This *is not* the general case, and the difference may be from few per cent to many times. There are two explanations of this: 1) a historical tradition to publish the coefficients in a form $C_\alpha \pm \sigma[C_\alpha]$, so other authors may restore only diagonal elements $R_{\alpha\alpha} = \sigma^2[C_\alpha]$, and not the non-diagonal ones; 2) "matrix – phobia", as described in presentations by Prof. Zdenek Mikulášek, who uses correct expressions ([20,21,22]).

The test function $\Phi_m$ is a particular case of $G(C_\alpha)$ and

$$\frac{\partial^2 \Phi_m}{\partial C_\alpha \partial C_\beta} = 2 A_{\alpha\beta} \qquad (22)$$

Combining (1) and (16), we obtain

$$\Phi_m(C_\alpha) = \Phi_{min} + \sum_{\alpha,\beta=1}^{m} A_{\alpha\beta} \delta C_\alpha \delta C_\beta \qquad (23)$$

$$\Phi_{min} = (\vec{x} - \overrightarrow{x_C}) \cdot (\vec{x} - \overrightarrow{x_C}) = (\vec{x}\vec{x}) - (\overrightarrow{x_C} \overrightarrow{x_C}) =$$
$$= \sum_{i,k=1}^{n} h_{ik} x_i x_k - \sum_{\alpha,\beta=1}^{m} A_{\alpha\beta} C_{0\alpha} C_{0\beta} \qquad (24)$$

Here $C_{0\alpha}$ is the solution of the system of normal equations (10) and $\delta C_\alpha$ are deviations of the coefficients $\delta C_\alpha$ from their "best fit"="statistically optimal" values $C_{0\alpha}$.

### 3. Determination of a statistically optimal number of coefficients

To determine the statistically optimal value of the number of parameters, the set of basic functions $f_\alpha(t), \alpha = 1 \ldots m_{max}$. From a sequence of test functions $\Phi_m$ for different $m = 1 \ldots m_{max}$, one has to determine the set of values

$$F_{q,n-m} = \frac{n-m}{q} \frac{\Phi_{m-q} - \Phi_m}{\Phi_m}, \qquad (23)$$

for $m = q+1 \ldots m_{max}$. Assuming that the observational errors obey the $n$ – dimensional Gaussian distribution, the parameter $F_{q,n-m}$ is a random variable which obeys the Fischer's distribution with $(q, n - m)$ degrees of freedom. The FAP (false alarm probability) is the probability that the random value exceeds the given number $F$:

$$\text{FAP} \stackrel{\text{def}}{=} \text{Prob}(\xi \geq F) = \int_F^\infty \rho_F(\xi)d\xi, \quad (24)$$

Typically one chooses the largest number of parameters, for which $\text{FAP} \leq \text{FAP}_{\text{crit}}$. The "critical value" $\text{FAP}_{\text{crit}}$ is chosen by the author and typically not greater than $10^{-2}$. Typically, $q = 1$ for the majority of cases, except adding periodic waves, when $q = 1$ parameters for a pair "sine and cosine" are added.

Another method for determination of $m$ is based on the minimization of accuracy at some specific points [3,4] or using the mean weighted value

$$\sigma_m^2[x_C] \stackrel{\text{def}}{=} \frac{1}{W}\sum_{k=1}^n w_k \, \sigma^2[x_C(t_k)], \quad (25)$$

$$W \stackrel{\text{def}}{=} \sum_{k=1}^n w_k \, \sigma^2[x_C(t_k)]. \quad (26)$$

Assuming Eq. (12), one gets

$$\sigma_m^2[x_C] \stackrel{\text{def}}{=} \frac{m\sigma_{0m}^2}{W}, \quad (27)$$

The minimum of this parameter may correspond to another estimate of the statistically optimal number of parameters $m$ [3,4]. There is no user – defined parameter needed, such as $\text{FAP}_{\text{crit}}$ for the Fischer's criterion, and thus is robustly defined. Typically this estimate of $m$ is significantly smaller than that for the Fischer's criterion with relatively large $\text{FAP}_{\text{crit}} \sim 10^{-2}$ or even smaller.

Other methods like finding the maximal value of the "signal/noise" ratio are discussed in [3,4].

## 2.2. Periodogram analysis of irregularly spaced arguments of signals

In the mathematical sense, the signal is periodic with a period $P$, if for all moments of time $t$ and integer numbers $k$ the signal satisfies equation

$$x(t + kP) = x(t). \quad (28)$$

Naturally, the value of the period $P$ should be the minimal positive one among all possible candidates, as all values $kP$ may be interpreted as "periods" (with $k$ equal parts).

In reality, one may have a number of pairs of observations $t_k, x_k$ (sometimes there may be a third parameter – the accuracy estimate $\sigma_k$ or weight $w_k$). It is suitable to define a "dimensionless" time $\psi = \frac{t - T_0}{P}$, where $T_0 -$ some moment of time, which is called "initial epoch". In astronomy, the initial epoch is typically set to some moment of maximum brightness of pulsating variable stars or minimum brightness of eclipsing binary stars [35].

To find hidden periodicities, there are many methods, which are based on the robust estimate of the "quality" of the "phase curve", i.e. the dependence of the signal not directly on time $t_k$, but on phase $\phi_k$, which is a function of time. For a constant period, one may write an expression $E + \phi = \psi$, $E -$ the (integer) cycle number, which is defined in a way that $0 \leq \phi < 1$ or $-0.5 \leq \phi < +0.5$.

The majority of methods are based on the "periodogram analysis", i.e. the dependence of the test function $S(f)$ on frequency $f = \frac{1}{P}$. This test function is computed over a grid of frequencies with a step $\Delta f = \frac{\Delta \phi}{t_n - t_1}$, where a recommended value of $\Delta \phi \approx 0.01 \dots 0.1$.

Then the peaks at the periodogram may correspond to periodicities in the signal. In some methods, the statistically optimal value of the period correspond to a minimum of $S(f)$ (some measure of scatter of points at the phase curve) or to a maximum.

The first group of methods may be called "point – point" or "non – parametric' periodogram analysis. It takes into account the weighted distance between the points. Different modifications are comparatively described in [4].

The second group of methods may be called "point – curve" or "parametric' periodogram analysis. In this case, the test function may be based on $\Phi_m(f)$. The (periodic) basic functions may be defined in an usual way: $f_1(\phi) = 1$, $f_{2j}(\phi) = \sin(2\pi j \phi)$, $f_{2j+1}(\phi) = \cos(2\pi j \phi)$, $j = 1 \dots s$ so the number of parameters for a fixed period is $m = 1 + 2s$.

This least squares method of the determination of the parameters in a case of (generally) irregularly spaced data is statistically correct, contrary to simplified formulae typically used for the Fourier transform, which is described e.g. in [1,17] and initially was applied to variable stars by Pickering (1881) [27] and Parenago and Kukarkin (1934) [26]. The complete method was applied to a gtoup of pulsating variables by [18]. The importance of the Fourier coefficients for the phase light curves of the eclipsing variable stars was studied by Kopal [16], Rucinski [31] et al.

For the periodogram analysis, it is suitable to define a dimensionless function

$$S(f) = r^2 = 1 - \frac{\Phi_m(f)}{\Phi_1}, \quad (29)$$

Because the first basic function is constant, $\Phi_1$ does not depend on frequency and is proportional to the variance of data. Here $r$ is the correlation coefficient between the observational $(x_k)$ and computed $(x_{Ck})$ values. If the data are normally distributed values, $S(f)$ is randomly distributed according to the Beta distribution with parameters $s$ and $(n - 1 - 2s)/2$. The FAP for the complete trigonometric polynomial of degree $s$ may be computed using the Fischer's distribution, as

$$F_{q,n-m} = \frac{n-m}{q}\frac{S}{1-S}$$
$$= \frac{n-1-2s}{2s}\frac{S}{1-S}. \quad (30)$$

However, this FAP is valid for the fixed frequency. However, the high peak may occur at one of many (typically up to millions) test frequencies. The computation of FAP and an estimate of the "effective number" of frequencies is discussed in [3,4].

Another problem is that the frequency corresponding to an extremum of $S(f)$ is determined on a grid, so the statistically optimal value is to be determined, e.g. using the method of differential corrections [3,4], adding an additional basic function

$$f_{2s+2}(t) = \varphi P \sum_{j=1}^{s} j(C_{2j}\cos(\varphi) - C_{2j+1}\sin(\varphi)), \quad (31)$$

where $\varphi = 2\pi(t - T_0)f$.

These "best fit" frequencies are dependent on a degree of the trigonometric polynomial $s$, so the number of parameters should be larger: $m = 2s + 2$.

To determine the statistically optimal value of $s$, one compares fits using trigonometric polynomials of orders $(s-1)$ and $s$, i.e. $q = 2$ in Eq. (23).

### 2.3. Special Shapes for Modeling Eclipses

Trigonometric polynomials are excellent tools for modeling periodic signals. However, Algol – type (EA) eclipsing binary stars show narrow deep eclipses, so one needs high order trigonometric polynomials. Thus seems natural to use "special shapes" – additional locally defined functions, which are equal to zero outside eclipses:

$$H(z) = \begin{cases} H(z), & \text{if } |z| \leq 1 \\ 0, & \text{if } |z| > 1 \end{cases} \quad (32)$$

Here $z = (\phi - \phi_0)/\Delta\phi$, $\phi_0$ is the phase of the center of the minimum, and $\Delta\phi$ is the filter half-width. The most simple function $H(z) = 1 - z^2$ for eclipses (and constant outside) was used earlier for automatic period determination and classification of new variable stars discovered from the space observatory Hipparcos.

Mikulašek et al. [20,21,22] used different modifications of the Gaussian $H(z) = \exp(-cz^2)$ which are suitable for approximation. The only problem is to determine the eclipse half-width, which is needed for the "General Catalogue of Variable Stars" [33] and other catalogs, as the Gaussian is infinite.

Andronov [5,6] used a local approximation

$$H(z,\alpha) = (1 - |z|^\alpha)^{1.5} \quad (33)$$

which has correct asymptotic behavior at $|z| \to 1$ as compared to a simplified physical model [2,9,10]. The shape of the eclipse is determined by a single parameter $\alpha$, which is generally different for the primary and secondary eclipses: the extreme case $\alpha \approx 1$ corresponds to a "triangle" shape of the minimum, which may be interpreted by stars with close radiuses; the "flat" minimum ($\alpha \gg 1$) corresponds to a total eclipse, when one (eclipsing) star significantly exceeds another one in size.

In the next section, we apply the methods described above, to concrete stars.

## 3. Applications to Eclipsing Binary Stars

### 3.1. The Observational Data

For the illustration of the methods described above, we have used the tables of the observations of the newly discovered Algol-type (EA) star 2MASS J11080308-6145589, published electronically by [23] and available in the Internet ( http://var.astro.cz/oejv/issues/oejv0102_appendix1.txt ).

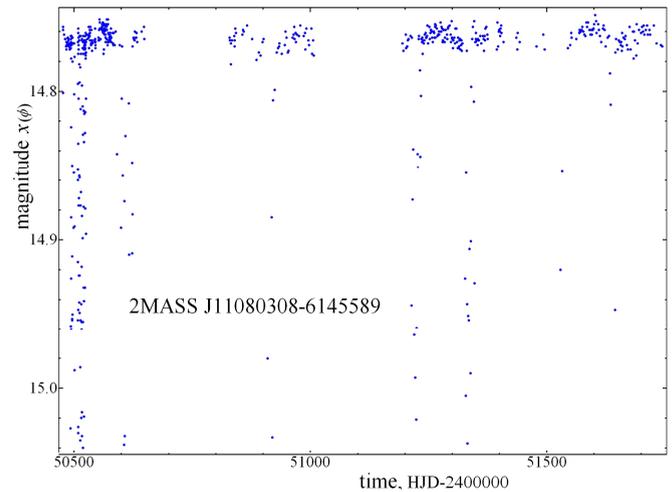

**Figure 1.** The time light curve of the eclipsing binary star 2MASS J11080308-6145589

### 3.2. Periodogram Analysis

The periodogram analysis was carried out using the test function $S(f)$ as defined in Eq. (29) and realized in the MCV program [8]. For the illustrative purposes, we used a set of the values of the degree of the trigonometric polynomial $s$ and the number of trial frequencies $f$ was set to 1 million for the duration of the observations. The data are the same as shown in Fig. 1.

The simplest periodogram for $s = 1$ shows a highest peak at $f = 1/1.006951 \approx 1$, while the true period $P$ corresponds to 2.013897, i.e. nearly twice larger than that obtained from the sine fit. The peaks at $f \approx 0$ and $f \approx 2$ cycles/day are called "daily aliases" or "daily bias" (e.g. [3,4]).

With an increasing $s$, the number of the apparent peaks increases drastically. The "true" peak increases in height when $s$ becomes even, and remains practically the same for odd $s$. The peak becomes higher than that at the double frequency $f \approx 1$ only for large $s \geq 14$.

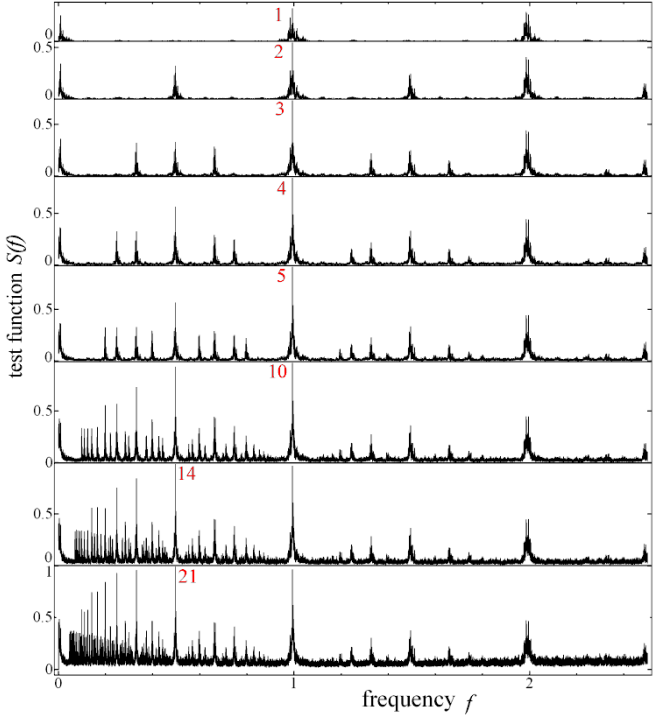

**Figure 2.** The periodograms $S(f)$ computed using the trigonometric polynomial approximations of different order $s$ (marked with a red color) 2MASS J11080308-6145589.

The light curve (dependence of the stellar magnitude $x$ on time $t$) is shown in Fig. 1. In astronomy, time is measured in days as the Heliocentric Julian Date (JD). The total duration of observations is 4 years, but the estimated period is close to 2 days, so at this scale, separate minima are not seen and the graph appears to be noisy.

### 3.3. Optimal Degree $s$ of the Trigonomerical Polynomial

The first criterion to determine the optimal number of parameters $m = 2 + 2s$ is the Fischer's criterion with $q = 2$ (Eq. (23)). In Figure 3, the dependence of $\gamma \stackrel{\text{def}}{=} -\lg$ FAP on $m$ is shown. One may see that, for odd $s$ (shown in green color), the parameter $\gamma$ is much smaller than that for the nearby even values of $s$. This is a typical situation for the eclipsing binary stars with comparable values of the depth of the minima. The statistically optimal number of parameters $m$ corresponding to the critical value of $\text{FAP}_{\text{crit}} = 10^{-2}$ is $m = 44$.

In Fig. 4 is shown another criterion $\sigma_m[x_C]$ (Eq. (27)) used for the determination of the optimal number of parameters. The green points (odd values of $s$) are higher than the preceding blue points (even values of $s$), as the increase of the number of corresponding parameters does not decrease significantly the shifts of the observations from the smoothing curve, but is more noisy due to a larger number of degrees of freedom. The minimum corresponds to $m = 40$, a number, which is smaller than that for the Fischer's criterion.

The NAV fit is characterized by much better mean accuracy of $\sigma_m[x_C] = 0.00109$, which is never reached by the TP fits even of high order. The r.m.s. deviation of the data from the NAV fit is 0.00742.

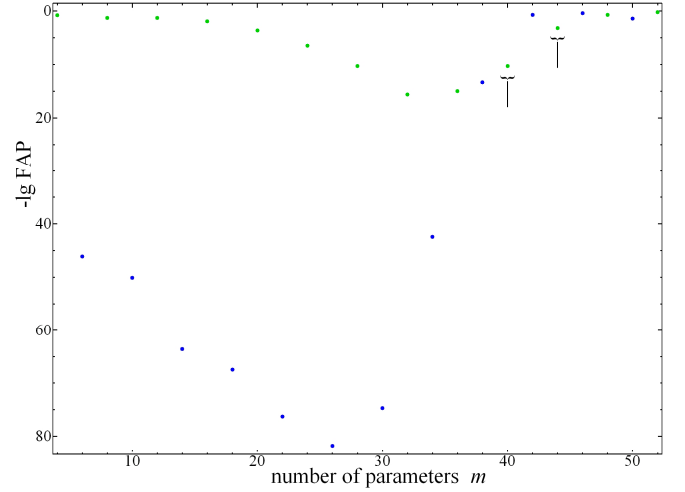

**Figure 3.** The dependence of the False Alarm Probability (FAP) on the number of parameters $m = 2s + 2$. The odd values of the degree of the trigonometric polynomial s are marked with a green color and the even s are marked with a blue color. The vertical bars correspond to the values $m = 40$ and $m = 44$.

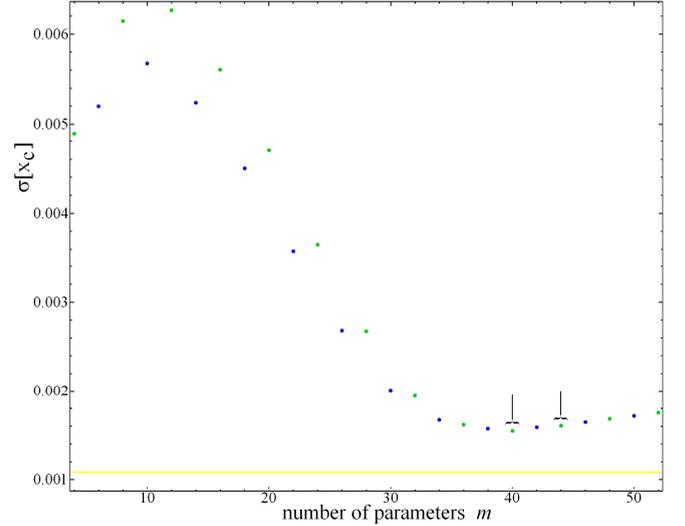

**Figure 4.** The dependence of the r.m.s. accuracy estimate of the smoothing function at the points of observations $\sigma_m[x_C]$ on the number of parameter $m = 2s + 2$. The odd values of the degree of the trigonometric polynomial s are marked with a green color and the even s are marked with a blue color. The vertical bars correspond to the values $m = 40$ and $m = 44$.

The coefficients $C_\alpha$ are shown in Fig. 5. One may note that they may be separated into few groups. The restoration of the smoothing function (5) only for these groups leads to the following conclusions: the sum of the terms in Eq. (5) for the red points only leads to "eclipses" occurring 4 times per period; the simultaneous sum with "blue" terms produces 2 eclipses per period, both of the similar depth; the different depth and shape of the eclipses is taken into account by the "green" terms (all of them correspond to the "cosine" basic functions), while the terms with the "sine" basic functions ("black") are responsible for the O'Connell effect [11,15,30,32,35], i.e. the asymmetry of the out-of-eclipse part of the phase light curve. The vertical bars show last pairs of coefficients, which are statistically significant according to two criteria mentioned above. Next coefficients are not statistically significant.

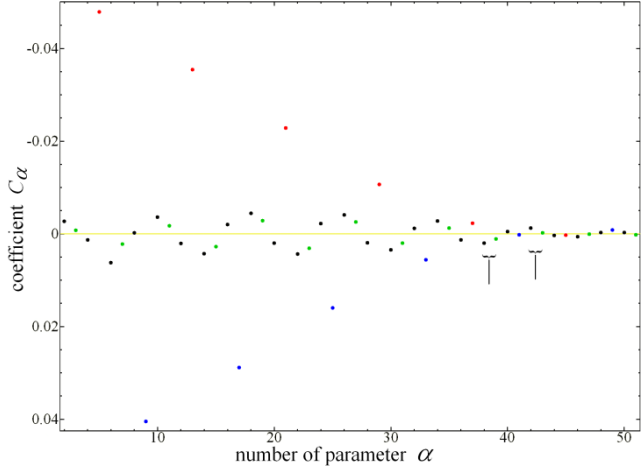

**Figure 5.** The dependence of the coefficients $C_\alpha$ of the smoothing trigonometric polynomial with $m = 50$ parameters. The odd values of the degree of the trigonometric polynomial s are marked with a green color and the even s are marked with red and blue colors. The black color marks coefficients for sine basic functions. The vertical bars correspond to the values $m = 40$ and $m = 44$. Yellow line shows the zero value.

### 3.4. The NAV Fit

The NAV fit contains 12 parameters, and the basic functions are:

$$f_1(\phi) = 1, f_2(\phi) = \cos(2\pi\phi), \; f_3(\phi) = \sin(2\pi\phi),$$
$$f_4(\phi) = \cos(4\pi\phi), f_5(\phi) = \sin(4\pi\phi),$$
$$f_6(\phi) = H\left(\frac{\phi}{C_8}; C_9\right), f_7(\phi) = H\left(\frac{\phi - 0.5}{C_8}; C_{10}\right), \quad (34)$$

$$\phi = \psi - \text{int}(\psi + 0.75),$$
$$\psi = \frac{t - T_0}{P_0} - (C_{11} + C_{12} \cdot (t - T_1))$$

Here $T_0$ – is the initial epoch and $P_0$ – is the initial period (e.g. determined from the same (or other) observations using any method). For our concrete star, we have used the values published by the discoverer of variability.

The general advice for the study is to determine the value of $T_0$ as that, which is most close in time to a (weighted) mean of the times of observations $\bar{t}$ (similar to [3]). In other words, we have to redefine

$$T_0 := T_0 + P_0 \cdot \text{int}\left(\frac{t - T_0}{P_0} + 0.5\right). \quad (35)$$

The published values [20] are $T_0 = 2450518.898$ and $P_0 = 2.0139$ (days). After the redefinition (Eq.(35)), we got $T_0 = 2450941.817$.

If we do not need the phase and/or period correction, the corresponding parameters $C_{11}$ and/or $C_{12}$ may be set to zero and should not be used in the differential corrections. For our illustrative case, we determined these parameters as well. The parameters are listed in Table 1 and the corresponding phase curve is shown in Fig. 6, where the 7 – parameter fit is shown, as well as the "abbreviated" 5 – parameter fit neglecting eclipses and showing (at the phases of both eclipses) an extrapolation of the "out-of-eclipse" part taking into account only the mean brightness ($C_1$), the reflection effect ($C_2$), the ellipticity effect ($C_4$), and the O'Connell effect ($C_3, C_5$). In fact, all these effects contribute to all parameters mentioned, but the major part of the amplitude corresponds to the parameters $C_\alpha$, as listed above.

**Table 1. Statistically Optimal Parameters of the NAV approximation**

| Parameter | Value | Parameter | Value |
|---|---|---|---|
| $C_1$ | 14.76486±0.00036 | $C_7$ | 0.18424±0.00257 |
| $C_2$ | -0.00229±0.00052 | $C_8$ | 0.04518±0.00043 |
| $C_3$ | -0.00105±0.00047 | $C_9$ | 1.40397±0.03697 |
| $C_4$ | 0.00470±0.00056 | $C_{10}$ | 1.90802±0.06658 |
| $C_5$ | -0.00003±0.00046 | $C_{11}$ | 0.00288±0.00014 |
| $C_6$ | 0.28127±0.00313 | $C_{12}$ | $(-8.47\pm3.24)\cdot 10^{-7}$ |

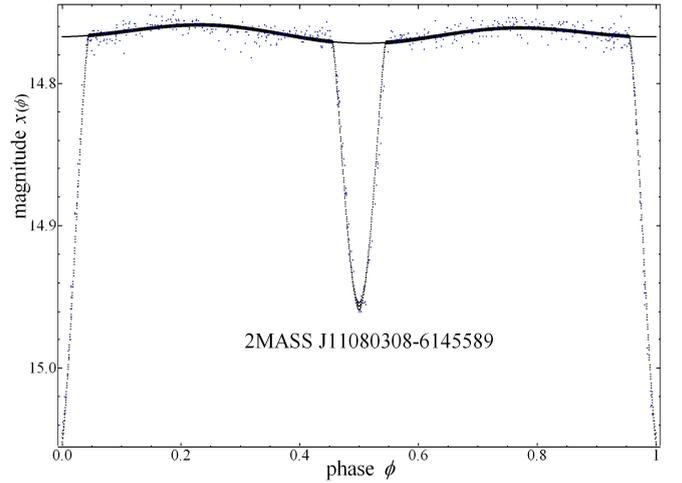

**Figure 6.** The phase light curve using the NAV approximation (Eq.(34)). The observations are shown as blue dots, the black dots show the approximation.

It should be noted that, for a given number of observations $n = 559$, the statistical error estimates relatively small. The coefficient $C_1$ corresponds to the mean brightness and thus is dependent not only on the system's luminosity, but also on the distance from the Earth. Other parameters are dependent either on the system characteristics, or on the orientation of the orbit (i.e. the inclination $i$, which is the angle between the rotational axis of the binary system, and the line of sight [2,11,30,31,35]). The O'Connell effect is not statistically significant in this system, the reflection effect is at the limit of detectability ($4.4\sigma$). However, the ellipticity effect is present at $8.4\sigma$ and thus is much more statistically significant.

The parameters $C_8 \ldots C_{12}$ were determined using the method of the differential corrections and are also statistically optimal.

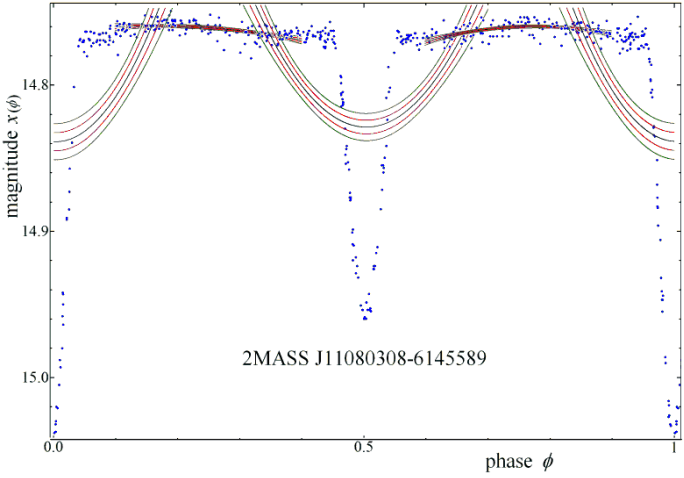

**Figure 7.** The phase light curve using the parabolic approximations in the phase intervals [-0.2,+0.2], [0.1,0.4], [0.3,0.7],[0.8,1.2] according to the method of [25] (blue lines). The red and green lines are $\pm 1\sigma$ and $\pm 2\sigma$ error corridors. The observations are shown as blue dots.

### 3.5. Comparison of the TP and NAV fits

In Fig. 8 is shown the dependence of the value of the smoothing function $x_C(\phi)$ at mid-eclipse (i.e. $\phi = 0$) on $m$. For small $m$, the value gradually decreases (again showing that the contributions of the waves with odd $s$ are much smaller than for even $s$. Finally they approach some horizontal line, but the value is shifted as compared to the NAV approximation. This means that, even for large $s$, there may be some systematic shifts between the NAV algorithm.

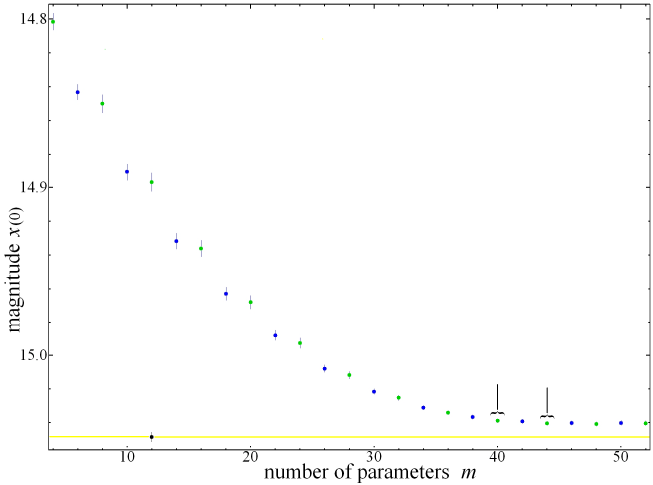

**Figure 8.** The dependence of the value of the smoothing function at phase zero (mid-eclipse) on the number of parameter $m = 2s + 2$. The odd values of the degree of the trigonometric polynomial s are marked with a green color and the even s are marked with a blue color. The vertical bars correspond to the values $m = 40$ and $m = 44$. Yellow line shows the value obtained by the NAV algorithm.

In Fig. 9 we show a small part of the light curve close to the bottom of the eclipse (phase $\phi \approx 0$) for two smoothing functions – the "New Algol Variable" (NAV) and the trigonometric polynomial of the statistically optimal degree $s = 21$ (TP21). Even for such a large value of $s$ (and the corresponding number of parameters $m = 44$), the TP21 fit still shows either absence of an "abrupt" minimum, or the "triangle – like" shape of the eclipse, as the curve is too smooth as compared to that for the NAV algorithm.

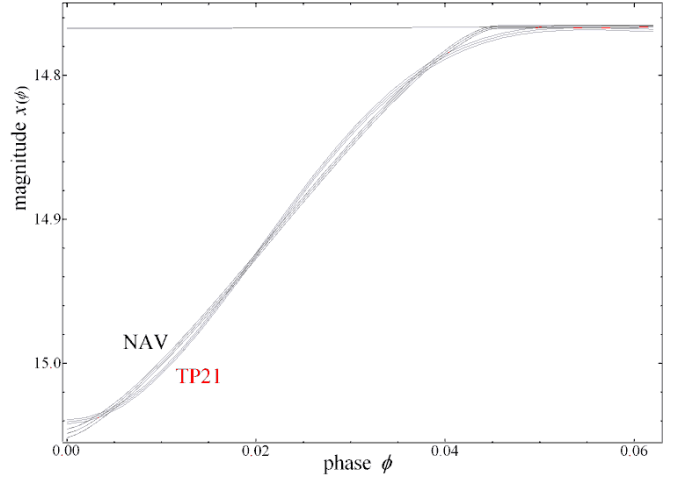

**Figure 9.** The approximations of the phase curve near the primary eclipse using the NAV and TP21 fits. For both approximations, the $\pm 1\sigma$ error corridors are shown.

### 3.6. The GCVS Parameters

The minimal set of the parameters needed for the GCVS [33] contains the co-ordinates (which have to be published by the discoverers of the variability of every star, so is the case for our object). Next pair of the parameters is the precession, which is computed analytically using the co-ordinates and thus is also out of the scope of the analysis of photometrical observations. The initial epoch and the period may be corrected using the NAV parameters $C_{11}$, $C_{12}$:

$$T_{01} = \frac{T_0 + C_{11}P_0 - P_0 C_{12} T_1}{1 - P_0 C_{12}} \quad (36)$$

$$P_1 = \frac{P_0}{1 - P_0 C_{12}} \quad (37)$$

It may be recommended to use the same redefined values From Eq.(35) for both $T_0$ and $T_1$. So, finally, $T_{01} = 2450941.82281 \pm 0.00028$ and $P_1 = 2.01389656 \pm 0.00000065$ days. In the discovery paper [23], the initial epoch was rounded to 0.001 days, so our accuracy estimate is 3.6 times better. Our accuracy estimate of the period $P_1$ is 61 (!) times better. This argues for a much better effectiveness of the NAV algorithm as compared to other methods.

The GCVS duration of the eclipse is measured in per cent of the orbital period (rounded to integers) and thus is related to the NAV parameter $C_8$:

$$D = \text{round}(200 C_8) \quad (38)$$

In the GCVS [33], the variability range is needed, but the magnitudes at the secondary minimum and secondary maximum are typically listed in the remarks. The brightness at the maximum is $m_{\text{MaxI}} = 14.7590 \pm 0.0008$, $m_{\text{minI}} = 15.0485 \pm 0.0030$, with a corresponding amplitude $\Delta m = 0.2896 \pm 0.0031$. The brightness at the secondary minimum is $m_{\text{minII}} = 14.9561 \pm 0.0024$ and at the secondary maximum $m_{\text{MaxII}} = 14.7611 \pm 0.0008$.

### 3.7. Additional Parameters

Phenomenological modeling is applied when there is no sufficient information to make a physical modeling. However, some physical constrains may be done in a simplified physical model [2, 10, 19, 34], where the stars are modeled as the spherical objects with a constant brightness over the limb.

As the eclipsed surface of the star (in the case of the circular orbit) is the same at the primary and the secondary minimum, one may introduce the parameters [5,6,9]

$$d_1 = 1 - 10^{-0.4C_6}, \quad (39)$$
$$d_2 = 1 - 10^{-0.4C_7}, \quad (40)$$
$$Y = d_1 + d_2, \quad (41)$$
$$\zeta = \frac{d_1}{d_2} = \frac{F_2}{F_1}, \quad (42)$$

Here $F_1$ and $F_2$ are mean surface brightness for the first and second star, respectively. Using adopted values of the parameters, we get $d_1 = 0.2282 \pm 0.0022$, $d_2 = 0.1561 \pm 0.0020$, $Y = 0.3843 \pm 0.0032$, $\zeta = 1.462 \pm 0.022$.

It is important to note that in previous papers (e.g. [19]) the authors had used a parameter $t = 10^{+0.4(m_{min1}-m_{max1})} - 1 = d_1/(1-d_1)$, where, instead of the corrected Eq. (39), they suggested $d_1 = 1 - 10^{-0.4(m_{min1}-m_{max1})}$.

### 3.8 W UMa-type stars show distinct eclipses?

Apparent advantages of the NAV algorithm are obvious for the EA - type systems, but need tests for the systems with smooth variations - EB and EW.

In Fig. 10, the best NAV fits for the phase curve of the EW – type system GSC 3692-0624 are shown for the observations published by Devlen [14]. The phase light curves in all 3 filters show distinct decrease of the amplitudes and depth of the primary minimum with the wavelength (i.e. in the sequence BVRc). Contrary, the depth of the secondary minimum increases with the wavelength. This argues that the temperatures of the components are significantly different, thus the system may *not* be in a thermal contact.

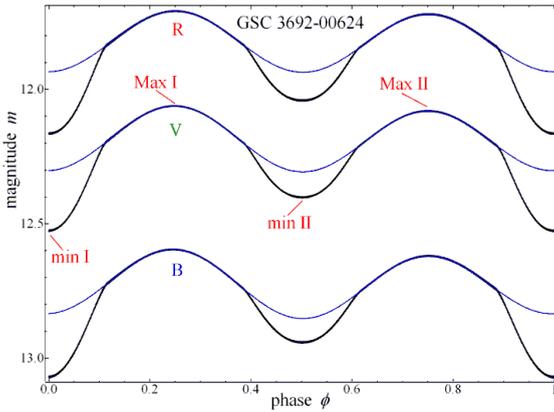

**Figure 10.** The phase light curve using the NAV approximation (Eq.(34)). The blue line corresponds to the 5-parameter restricted, the black line shows the complete 7-parameter fit.

Anyway, the $H(z)$ shapes are present even with a large amplitudes of the TP2 shape of the out-of-eclipse part of the curve. This contradicts the statement in the majority of classification schemes in the textbooks that "the start and end of eclipses are seen only in the Algol-type (EA) systems, but not in EB and EW". Our experience shows that zero-values of the eclipse depth (within few sigma) correspond to small amplitude of the out-of eclipse variability, and thus the low inclination i, and eclipses may not be present, arguing for elliptical variability and the "Ell" classification.

## 4. Conclusions and Results

In this paper, we compared classical trigonometric polynomial (TP) algorithm with the "New Algol Variable" (NAV) one. Although both are "phenomenological" and approximate observations of eclipsing variables, the NAV algorithm has few advantages:

1. The number of parameters needed to obtain the approximation of the same quality (in the sense of the Least Squares) is much smaller than for the TP fits. The effectiveness o of the eclipse of the NAV method increases with decreasing width of the eclipse.

2. Contrary to the TP, the NAV approximation allows determination of the important parameter needed for the official registration in the "General Catalogue of Variable Stars" - the phase width of the eclipse.

3. With the NAV algorithm, the definition of the depth of the eclipse may be redefined - instead of the difference of brightness between the minimum and maximum $\Delta m$, we introduce the parameters $C_6, C_7$, , which remove the influence onto $\Delta m$ of the possible effects of O'Connell, reflection and ellipticity. This leads to more correct estimate of the surface brightness ratio $\zeta$.

4. The parameters $C_9, C_{10}$, which describe the shape of the eclipses, are expected to be the same in the model of constant surface brightness for each star (neglecting the limb darkening). However, in reality they are typically different, larger value of this parameter corresponds to a larger radius in a binary system.

5. The parameters $C_{11}, C_{12}$ are introduced for the case of the possible phase shift of the minimum from zero and of the deviation of the best fit period from its preliminary determined value $P_0$.

6. For one-color photometrical observations, one may determine more accurate values of the depth of the minima, which may give information on the ratio of the surface brightness of the eclipsing stars, which may be then used along with the statistical "mass-radius-temperature dependencies" for the stars. The parameter shows the degree of eclipses and thus one may distinguish systems without eclipses $\zeta = 0$ and with both full eclipses $\zeta = 1$. If there is no full eclipse in the system, the analysis is more complicated, but possible with using an additional information.

# Acknowledgements

ILA thanks Professors Yonggi Kim, Valentin G. Karetnikov, Zdenek Mikulášek, David Turner and Valery Yu. Terebizh for fruitful discussions on eclipsing binaries and different methods of the time series analysis. This work is a part of the "Inter-longitude Astronomy" [8] and "Ukrainian Virtual Observatory" [36] projects.